\documentclass[journal,a4paper,12pt,onecolumn]{IEEEtran}
\usepackage{amssymb}
\usepackage{amsmath}
\usepackage{amsfonts}

\begin{document}

\title{On Controllability and Near-controllability of Multi-input
Discrete-time Bilinear Systems in Dimension Two*\thanks{%
*This work was supported by the China Postdoctoral Science Foundation funded
project and the National Natural Science Foundation of China.}}
\author{Lin Tie$^{\dag }$\thanks{$^{\dag }$The author is with the School of
Automation Science and Electrical Engineering, Beihang University (Beijing
University of Aeronautics and Astronautics), Beijing, P. R. China. Email:
\texttt{\small tielinllc@gmail.com}}}
\maketitle

\begin{abstract}
This paper completely solves the controllability problems of two-dimensional
multi-input discrete-time bilinear systems with and without drift. Necessary
and sufficient conditions for controllability, which cover the existing
results, are obtained by using an algebraic method. Furthermore, for the
uncontrollable systems, near-controllability is studied and necessary and
sufficient conditions for the systems to be nearly controllable are also
presented. Examples are provided to demonstrate the conceptions and results
of the paper.
\end{abstract}


\section{Introduction}

$\left. {}\right. $

\setcounter{page}{1}Bilinear systems form an important class of nonlinear
systems which have attracted a great deal of attention over the past decades
[1]. The main interest of such systems lies in the fact that they are not
only good models to represent a large classes of real-world processes
ranging from engineering to non-engineering fields (e.g. chemistry, biology,
and socio-economics [2-5]), but also simpler and better understood than most
other nonlinear systems. It is reasonable to say that bilinear systems have
nowadays been one of the focuses in the literature of nonlinear systems.

$\left. {}\right. $

Controllability is clearly an important issue in mathematical control
theory. The controllability problems of bilinear systems were raised at the
beginning of the study on such systems [2]. Today, controllability has
become one of the hot topics in bilinear systems. This is particularly true
for the continuous-time case. More specifically, controllability of
continuous-time bilinear systems has been well studied profiting from the
Lie algebra methods, and various Lie-algebraic criterions have been obtained
in the literature [1]. For discrete-time bilinear systems, however,
available results on controllability are very sparse. In particular, most of
the existing works on controllability of discrete-time bilinear systems
focus on the single-input case [5-12], while for the multi-input case, few
work has been reported except for [13-15].

$\left. {}\right. $

The multi-input discrete-time bilinear systems can be described by the
following difference equation%
\begin{eqnarray}
x\left( k+1\right)  &=&Ax\left( k\right) +u_{1}\left( k\right) B_{1}x\left(
k\right) +\cdots +u_{m}\left( k\right) B_{m}x\left( k\right)   \notag \\
&=&\left( A+u_{1}\left( k\right) B_{1}+\cdots +u_{m}\left( k\right)
B_{m}\right) x\left( k\right)   \notag \\
&=&\left( A+\underset{i=1}{\overset{m}{\sum }}u_{i}\left( k\right)
B_{i}\right) x\left( k\right)
\end{eqnarray}%
where $x(k)\in
\mathbb{R}
^{n}$, $A,B_{1},\ldots ,B_{m}\in
\mathbb{R}
^{n\times n}$, $u_{1}(k),\ldots ,u_{m}\left( k\right) \in
\mathbb{R}
$, and $m\geq 2$. In particular, $Ax\left( k\right) $\ is called the drift
term. If $A$ can be linearly represented by $B_{1},\ldots ,B_{m}$, then
system (1) degenerates to the multi-input discrete-time bilinear system
without drift%
\begin{eqnarray}
x\left( k+1\right)  &=&u_{1}\left( k\right) B_{1}x\left( k\right) +\cdots
+u_{m}\left( k\right) B_{m}x\left( k\right)   \notag \\
&=&\left( u_{1}\left( k\right) B_{1}+\cdots +u_{m}\left( k\right)
B_{m}\right) x\left( k\right)   \notag \\
&=&\underset{i=1}{\overset{m}{\sum }}u_{i}\left( k\right) B_{i}x\left(
k\right)
\end{eqnarray}%
where $x(k)\in
\mathbb{R}
^{n}$, $B_{1},\ldots ,B_{m}\in
\mathbb{R}
^{n\times n}$, $u_{1}(k),\ldots ,u_{m}\left( k\right) \in
\mathbb{R}
$, and $m\geq 2$. Therefore, without loss of generality, it is assumed
throughout this paper that, for system (1), $A,B_{1},\ldots ,B_{m}$ are
linearly independent and that, for system (2), $B_{1},\ldots ,B_{m}$ are
linearly independent. For controllability of system (1), there are only
scarce results [13-15]. [13] proposed sufficient conditions (mainly for the
case of $m=2$) under the two strong assumptions that all of $B_{1},\ldots
,B_{m}$ have rank one and only one of $u_{1}(k),\ldots ,u_{m}\left( k\right)
$ is applied at any one instant. [14,15] considered system (1) in dimension
two with $A$ a scalar matrix and obtained a necessary and sufficient
condition based on the controllability properties of the corresponding
single-input system. For system (2), its controllability is unknown. It
should be noted that, for controllability of both single-input and
multi-input discrete-time bilinear systems, only special subclasses have
been considered, while most cases remain unsolved. Even for the systems (no
matter single-input or multi-input) in dimension two, there does not exist a
general necessary and sufficient condition for controllability [12].

$\left. {}\right. $

In this paper, we still focus on the multi-input systems in dimension two
and study the controllability problems by using an algebraic and
straightforward method without considering the single-input system.
Necessary and sufficient conditions for two-dimensional systems (1) and (2)
to be controllable are presented, which cover the results obtained in
[13-15] and are easy to apply. Furthermore, the algebraic method also makes
it possible to get the required control inputs which achieve the state
transition for the controllable systems. Then, for the uncontrollable
systems (1) and (2) which may have a large controllable region\footnote{%
A controllable region is such a region that, for any two states in the
region, the transition from one to the other can be achieved by admissible
controls.}, the near-controllability problems are addressed and necessary
and sufficient conditions for the systems to be nearly controllable are
given. As a result, the controllability problems of multi-input
discrete-time bilinear systems in dimension two are completely solved and
the controllability properties of such systems are fully characterized.
Finally, examples are provided to demonstrate the results of the paper.

$\left. {}\right. $

This paper is organized as follows. Section II is devoted to controllability
of systems (1) and (2), while Section III is devoted to near-controllability
of systems (1) and (2). Examples are shown in Section IV and concluding
remarks of the paper are made in Section V.

$\left. {}\right. $

\section{Controllability}

$\left. {}\right. $

We first give the controllability definition of systems (1) and (2).

$\left. {}\right. $

\textbf{Definition 1.} Systems (1) and (2) are said to be controllable if,
for any $\xi ,\eta $ in$\
\mathbb{R}
_{\ast }^{n}$ $\left(
\mathbb{R}
_{\ast }^{n}:=%
\mathbb{R}
^{n}\left\backslash \left\{ 0\right\} \right. \right) $, there exist a
positive integer $l$\ and a finite control sequence $\left( u_{1}(0),\ldots
,u_{m}\left( 0\right) \right) ,\left( u_{1}(1),\ldots ,u_{m}\left( 1\right)
\right) ,$ $\ldots ,\left( u_{1}(l-1),\ldots ,u_{m}\left( l-1\right) \right)
$ such that $\xi $\ can be transferred to $\eta $ at step $k=l$.

$\left. {}\right. $

To prove controllability of system (1) in dimension two, we need the
following lemmas.

$\left. {}\right. $

\textbf{Lemma 1.} If $B_{1},B_{2}\in
\mathbb{R}
^{2\times 2}$ are linearly independent and do not have a real eigenvector in
common, then the set%
\begin{equation}
\left\{ \left. \zeta =\left[
\begin{array}{cc}
\zeta _{1} & \zeta _{2}%
\end{array}%
\right] ^{T}\in
\mathbb{R}
^{2}\text{ }\right\vert \text{ }\left\vert
\begin{array}{cc}
B_{1}\zeta & B_{2}\zeta%
\end{array}%
\right\vert =0\right\}
\end{equation}%
is a point $\left\{ 0\right\} $ or forms one line in $%
\mathbb{R}
^{2}$ or two lines in $%
\mathbb{R}
^{2}$, where $\left\vert \cdot \right\vert $ denotes the determinant of a
matrix throughout this paper.

$\left. {}\right. $

\textbf{Proof.} Note that $\left\vert
\begin{array}{cc}
B_{1}\zeta & B_{2}\zeta%
\end{array}%
\right\vert $ is either identically equal to zero or a second-degree
homogeneous polynomial with respect to $\zeta _{1},\zeta _{2}$. Write%
\begin{equation*}
B_{1}=\left[
\begin{array}{cc}
\alpha _{1} & \beta _{1}%
\end{array}%
\right] ,\text{ }B_{2}=\left[
\begin{array}{cc}
\alpha _{2} & \beta _{2}%
\end{array}%
\right]
\end{equation*}%
where $\alpha _{1},\beta _{1},\alpha _{2},\beta _{2}\in
\mathbb{R}
^{2}$. If%
\begin{equation}
\left\vert
\begin{array}{cc}
B_{1}\zeta & B_{2}\zeta%
\end{array}%
\right\vert \equiv 0
\end{equation}%
for any $\zeta \in
\mathbb{R}
^{2}$, let $\zeta =\left[
\begin{array}{cc}
1 & 0%
\end{array}%
\right] ^{T},\left[
\begin{array}{cc}
0 & 1%
\end{array}%
\right] ^{T}$ respectively and from (4) we obtain that $\alpha _{1},\alpha
_{2}$ are linearly dependent and $\beta _{1},\beta _{2}$ are linearly
dependent. Since $B_{1},B_{2}$ are linearly independent, at least one of $%
\alpha _{1},\alpha _{2}$ is nonzero. Assume that it is $\alpha _{1}$ and we
can write $\alpha _{2}=p\alpha _{1}$ for some $p\in
\mathbb{R}
$. In addition, by noting%
\begin{eqnarray*}
\left\vert
\begin{array}{cc}
B_{1}\zeta & B_{2}\zeta%
\end{array}%
\right\vert &\equiv &\left\vert
\begin{array}{cc}
B_{1}\zeta & \left( B_{2}-pB_{1}\right) \zeta%
\end{array}%
\right\vert \\
&=&\left\vert
\begin{array}{cc}
B_{1}\zeta & \left[
\begin{array}{cc}
0 & \beta _{2}-p\beta _{1}%
\end{array}%
\right] \zeta%
\end{array}%
\right\vert ,
\end{eqnarray*}%
we can further assume that $\alpha _{2}=0$ (i.e. $p=0$). Then $\beta
_{2}\neq 0$ (otherwise $B_{2}=0$ and $B_{1},B_{2}$\ are linearly
independent) and by a similar analysis we can assume that $\beta _{1}=0$.
Now let $\xi =\left[
\begin{array}{cc}
1 & 1%
\end{array}%
\right] ^{T}$ and from (4)%
\begin{equation*}
\left\vert
\begin{array}{cc}
B_{1}\zeta & B_{2}\zeta%
\end{array}%
\right\vert =\left\vert
\begin{array}{cc}
\alpha _{1} & \beta _{2}%
\end{array}%
\right\vert =0,
\end{equation*}%
which implies that $\alpha _{1},\beta _{2}$\ are linearly dependent. Write $%
\beta _{2}=r\alpha _{1}$ for some $r\in
\mathbb{R}
_{\ast }$. Then one can get that $\alpha _{1}$ is a common eigenvector of $%
B_{1},B_{2}$, which contradicts that $B_{1},B_{2}$ do not have a real
eigenvector in common.

Therefore, $\left\vert
\begin{array}{cc}
B_{1}\zeta & B_{2}\zeta%
\end{array}%
\right\vert $ is a second-degree homogeneous polynomial with respect to $%
\zeta _{1},\zeta _{2}$ and can be written as%
\begin{equation*}
\left\vert
\begin{array}{cc}
B_{1}\zeta & B_{2}\zeta%
\end{array}%
\right\vert =a\zeta _{1}^{2}+b\zeta _{1}\zeta _{2}+c\zeta _{2}^{2}
\end{equation*}%
where $a,b,c\in
\mathbb{R}
$ are related to $B_{1},B_{2}$. As a result, if $\left\vert
\begin{array}{cc}
B_{1}\zeta & B_{2}\zeta%
\end{array}%
\right\vert $ is a positive or negative polynomial then set (3) contains
only one point $\left\{ 0\right\} $ (this case corresponds to $b^{2}-4ac<0$%
). Otherwise, set (3) forms one line in $%
\mathbb{R}
^{2}$ (this case corresponds to $b^{2}-4ac=0$) or two lines in $%
\mathbb{R}
^{2}$ (this case corresponds to $b^{2}-4ac>0$). \ $\square $

$\left. {}\right. $

\textbf{Lemma 2.} If $B_{1},B_{2}\in
\mathbb{R}
^{2\times 2}$ are linearly independent and have a real eigenvector in
common, then the set as given in (3) is equal to $%
\mathbb{R}
^{2}$ only if there exists a nonsingular $P\in
\mathbb{R}
^{2\times 2}$ such that%
\begin{equation}
PB_{1}P^{-1}=\left[
\begin{array}{cc}
B_{11}^{1} & B_{12}^{1} \\
0 & 0%
\end{array}%
\right] ,\text{ }PB_{2}P^{-1}=\left[
\begin{array}{cc}
B_{11}^{2} & B_{12}^{2} \\
0 & 0%
\end{array}%
\right] .
\end{equation}%
Otherwise, set (3) forms one line in $%
\mathbb{R}
^{2}$ or two lines in $%
\mathbb{R}
^{2}$.

$\left. {}\right. $

\textbf{Proof.} Since $B_{1},B_{2}$ have a real eigenvector in common (named
$\alpha $), it can be easily verified that $PB_{1}P^{-1},PB_{2}P^{-1}$ are
both upper-triangular, where $P=\left[
\begin{array}{cc}
\alpha & \beta%
\end{array}%
\right] ^{-1}$ with $\beta $ any vector in $%
\mathbb{R}
^{2}$ linearly independent on $\alpha $. As a result, we can write%
\begin{equation*}
B_{1}=\left[
\begin{array}{cc}
B_{11}^{1} & B_{12}^{1} \\
0 & B_{22}^{1}%
\end{array}%
\right] ,\text{ }B_{2}=\left[
\begin{array}{cc}
B_{11}^{2} & B_{12}^{2} \\
0 & B_{22}^{2}%
\end{array}%
\right]
\end{equation*}%
without loss of generality. If%
\begin{equation*}
\left\vert
\begin{array}{cc}
B_{1}\zeta & B_{2}\zeta%
\end{array}%
\right\vert \equiv 0
\end{equation*}%
for any $\zeta $ in $%
\mathbb{R}
^{2}$, by a similar analysis as shown in the proof of Lemma 1 we can deduce $%
B_{22}^{1}=0,B_{22}^{2}=0$. Otherwise, $\left\vert
\begin{array}{cc}
B_{1}\zeta & B_{2}\zeta%
\end{array}%
\right\vert $ is a second-degree homogeneous polynomial with respect to $%
\zeta _{1},\zeta _{2}$. Note that set (3) always includes span$\left\{ \left[
\begin{array}{cc}
1 & 0%
\end{array}%
\right] ^{T}\right\} $. The set forms one line in $%
\mathbb{R}
^{2}$ or two lines in $%
\mathbb{R}
^{2}$.$\ \ \square $

$\left. {}\right. $

\textbf{Lemma 3.} If $A,B_{1},B_{2}\in
\mathbb{R}
^{2\times 2}$ are linearly independent and do not have a real eigenvector in
common, then, for any nonzero $\xi $ in%
\begin{equation}
\text{span}\left\{ \alpha \right\} \cup \text{span}\left\{ \beta \right\}
\end{equation}%
where $\alpha ,\beta \in
\mathbb{R}
_{\ast }^{2}$ are linearly independent, there exist $a,b\in
\mathbb{R}
$ such that $\left( A+aB_{1}+bB_{2}\right) \xi $ does not belong to (6).

$\left. {}\right. $

The proof of the above lemma is a little long and is hence put in appendix.

$\left. {}\right. $

\textbf{Lemma 4.} The system%
\begin{equation}
x\left( k+1\right) =\left( \left[
\begin{array}{cc}
0 & 0 \\
A_{21} & A_{22}%
\end{array}%
\right] +u_{1}\left( k\right) \left[
\begin{array}{cc}
1 & 0 \\
0 & 0%
\end{array}%
\right] +u_{2}\left( k\right) \left[
\begin{array}{cc}
0 & 1 \\
0 & 0%
\end{array}%
\right] \right) x\left( k\right)
\end{equation}%
is controllable, where $x(k)\in
\mathbb{R}
^{2}$, $u_{1}(k),u_{2}\left( k\right) \in
\mathbb{R}
$, and $A_{21}\neq 0$.

$\left. {}\right. $

\textbf{Proof.} Given $\xi =\left[
\begin{array}{cc}
\xi _{1} & \xi _{2}%
\end{array}%
\right] ^{T},\eta =\left[
\begin{array}{cc}
\eta _{1} & \eta _{2}%
\end{array}%
\right] ^{T}$ in $%
\mathbb{R}
_{\ast }^{2}$. If $\xi _{1}=0$, then $\xi _{2}\neq 0$ and letting $%
u_{1}\left( 0\right) =0,u_{2}\left( 0\right) $ be a sufficiently large
number we can transfer $\xi $ to a state $\zeta $ with not only $\zeta
_{1}\neq 0$ but also $A_{21}\zeta _{1}+A_{22}\zeta _{2}\neq 0$ since $%
A_{21}\neq 0$. Thus, for $\xi $,\ we can assume $\xi _{1}\neq 0$ and $%
A_{21}\xi _{1}+A_{22}\xi _{2}\neq 0$ without loss of generality. Now, let $%
u_{2}\left( 0\right) =0$. From (7)%
\begin{equation}
\left[
\begin{array}{cc}
u_{1}\left( 1\right)  & u_{2}\left( 1\right)  \\
A_{21} & A_{22}%
\end{array}%
\right] \left[
\begin{array}{cc}
u_{1}\left( 0\right)  & 0 \\
A_{21} & A_{22}%
\end{array}%
\right] \left[
\begin{array}{c}
\xi _{1} \\
\xi _{2}%
\end{array}%
\right] =\left[
\begin{array}{c}
u_{1}\left( 1\right) u_{1}\left( 0\right) \xi _{1}+u_{2}\left( 1\right)
\left( A_{21}\xi _{1}+A_{22}\xi _{2}\right)  \\
A_{21}\xi _{1}u_{1}\left( 0\right) +A_{22}\left( A_{21}\xi _{1}+A_{22}\xi
_{2}\right)
\end{array}%
\right] .  \notag
\end{equation}%
It can be verified that by%
\begin{eqnarray*}
u_{1}\left( 0\right)  &=&\frac{\eta _{2}-A_{22}\left( A_{21}\xi
_{1}+A_{22}\xi _{2}\right) }{A_{21}\xi _{1}},\text{ }u_{2}\left( 0\right) =0,
\\
u_{1}\left( 1\right)  &=&\frac{A_{21}\eta _{1}}{\eta _{2}-A_{22}\left(
A_{21}\xi _{1}+A_{22}\xi _{2}\right) },\text{ }u_{2}\left( 1\right) =0
\end{eqnarray*}%
if $\eta _{2}-A_{22}\left( A_{21}\xi _{1}+A_{22}\xi _{2}\right) \neq 0$ and%
\begin{eqnarray*}
u_{1}\left( 0\right)  &=&0,\text{ }u_{2}\left( 0\right) =0, \\
u_{1}\left( 1\right)  &=&0,\text{ }u_{2}\left( 1\right) =\frac{\eta _{1}}{%
A_{21}\xi _{1}+A_{22}\xi _{2}}
\end{eqnarray*}%
if $\eta _{2}-A_{22}\left( A_{21}\xi _{1}+A_{22}\xi _{2}\right) =0$, $\xi $
is transferred to $\eta $. \ $\square $

$\left. {}\right. $

\textbf{Lemma 5.} If $A,B_{1},B_{2},B_{3}\in
\mathbb{R}
^{2\times 2}$ do not have a real eigenvector in common, then there exists a
linear combination of $B_{2},B_{3}$, named $\left( \tilde{a}B_{2}+\tilde{b}%
B_{3}\right) $, such that $A,B_{1},\left( \tilde{a}B_{2}+\tilde{b}%
B_{3}\right) $ do not have a real eigenvector in common.

$\left. {}\right. $

\textbf{Proof.\ }If for any $\tilde{a},\tilde{b}\in
\mathbb{R}
$, $A,B_{1},\left( \tilde{a}B_{2}+\tilde{b}B_{3}\right) $ will have one
common real eigenvector, then by letting $\tilde{a}=1,\tilde{b}=0$ and $%
\tilde{a}=0,\tilde{b}=1$ respectively, it follows that $A,B_{1},B_{2}$ have
one common real eigenvector (named $\alpha $) and $A,B_{1},B_{3}$ have one
common real eigenvector (named $\beta $). Now let $\tilde{a}=1,\tilde{b}=1$.
$A,B_{1},\left( B_{2}+B_{3}\right) $ have one common real eigenvector which
must be either of $\alpha ,\beta $ as a $2\times 2$ matrix has at most two
real eigenvectors. Assume that it is $\alpha $. Then%
\begin{equation*}
\left( B_{2}+B_{3}\right) \alpha =\tilde{b}_{1}\alpha =B_{2}\alpha
+B_{3}\alpha =\tilde{b}_{2}\alpha +B_{3}\alpha \Rightarrow B_{3}\alpha
=\left( \tilde{b}_{1}-\tilde{b}_{2}\right) \alpha
\end{equation*}%
for some $\tilde{b}_{1},\tilde{b}_{2}\in
\mathbb{R}
$, which implies that $\alpha $ is also an eigenvector of $B_{3}$. However,
this contradicts that $A,B_{1},B_{2},B_{3}$ do not have a real eigenvector
in common. \ $\square $

$\left. {}\right. $

With the help of Lemmas 1-5, we can obtain the following conclusion on
controllability of system (1).

$\left. {}\right. $

\textbf{Theorem 1.} Consider system (1) with $n=2$. The system is
controllable if and only if $A,B_{1},\ldots ,B_{m}$ do not have a real
eigenvector in common.

$\left. {}\right. $

\textbf{Proof.} For the sufficiency, since $n=2$ and $A,B_{1},\ldots ,B_{m}$
are linearly independent, we have $m=2$ or $m=3$. First consider the case of
$m=2$. Given initial and terminal states $\xi ,\eta $ in $%
\mathbb{R}
_{\ast }^{2}$. If%
\begin{equation}
\left\vert
\begin{array}{cc}
B_{1}\xi  & B_{2}\xi
\end{array}%
\right\vert \neq 0,
\end{equation}%
then%
\begin{equation*}
\left[
\begin{array}{c}
u_{1}\left( 0\right)  \\
u_{2}\left( 0\right)
\end{array}%
\right] =\left[
\begin{array}{cc}
B_{1}\xi  & B_{2}\xi
\end{array}%
\right] ^{-1}\left( \eta -A\xi \right)
\end{equation*}%
are the control inputs that steer the system from $\xi $ to $\eta $. If, for
any $\xi $ in $%
\mathbb{R}
_{\ast }^{2}$, condition (8) holds, then we are through. Otherwise, if $%
B_{1},B_{2}$ cannot be simultaneously transformed into the forms as given in
(5), then, by Lemmas 1 and 2, the set%
\begin{equation}
\left\{ \left. \zeta \in
\mathbb{R}
^{2}\text{ }\right\vert \text{ }\left\vert
\begin{array}{cc}
B_{1}\zeta  & B_{2}\zeta
\end{array}%
\right\vert =0\right\}
\end{equation}%
at most forms one line in $%
\mathbb{R}
^{2}$ or two lines in $%
\mathbb{R}
^{2}$. By Lemma 3, for any $\xi $ in (9), we can find $u_{1}\left( 0\right)
=a,u_{2}\left( 0\right) =b$ to steer the system from $\xi $ to a state $\bar{%
\xi}$\ which is out of (9). Then, by%
\begin{equation*}
\left[
\begin{array}{c}
u_{1}\left( 1\right)  \\
u_{2}\left( 1\right)
\end{array}%
\right] =\left[
\begin{array}{cc}
B_{1}\bar{\xi} & B_{2}\bar{\xi}%
\end{array}%
\right] ^{-1}\left( \eta -A\bar{\xi}\right)
\end{equation*}%
the system is steered from $\bar{\xi}$ to $\eta $.

If $B_{1},B_{2}$ can be transformed into the forms as given in (5), we may
assume $B_{11}^{1}\neq 0,B_{12}^{2}\neq 0$ without loss of generality since $%
B_{1},B_{2}$ are linearly independent. Write%
\begin{equation*}
A=\left[
\begin{array}{cc}
A_{11} & A_{12} \\
A_{21} & A_{22}%
\end{array}%
\right] .
\end{equation*}%
We have $A_{21}\neq 0$ as $A,B_{1},B_{2}$ do not have a real eigenvector in
common. Denote%
\begin{equation*}
\left[
\begin{array}{c}
\bar{u}_{1}\left( k\right)  \\
\bar{u}_{2}\left( k\right)
\end{array}%
\right] =\left[
\begin{array}{cc}
B_{11}^{1} & B_{11}^{2} \\
B_{12}^{1} & B_{12}^{2}%
\end{array}%
\right] \left[
\begin{array}{c}
u_{1}\left( k\right)  \\
u_{2}\left( k\right)
\end{array}%
\right] +\left[
\begin{array}{c}
A_{11} \\
A_{12}%
\end{array}%
\right]
\end{equation*}%
where%
\begin{equation*}
\left[
\begin{array}{cc}
B_{11}^{1} & B_{11}^{2} \\
B_{12}^{1} & B_{12}^{2}%
\end{array}%
\right]
\end{equation*}%
is nonsingular since $B_{1},B_{2}$ are linearly independent. Then, the
system can be rewritten as%
\begin{equation*}
x\left( k+1\right) =\left( \left[
\begin{array}{cc}
0 & 0 \\
A_{21} & A_{22}%
\end{array}%
\right] +\bar{u}_{1}\left( k\right) \left[
\begin{array}{cc}
1 & 0 \\
0 & 0%
\end{array}%
\right] +\bar{u}_{2}\left( k\right) \left[
\begin{array}{cc}
0 & 1 \\
0 & 0%
\end{array}%
\right] \right) x\left( k\right)
\end{equation*}%
which is controllable according to Lemma 4. Controllability of the case of $%
m=2$ is thus proved.

For the case of $m=3$, by Lemma 5 we can have $\tilde{a},\tilde{b}\in
\mathbb{R}
$ such that $A,B_{1},\left( \tilde{a}B_{2}+\tilde{b}B_{3}\right) $ do not
have a real eigenvector in common. Clearly, $A,B_{1},\left( \tilde{a}B_{2}+%
\tilde{b}B_{3}\right) $ are also linearly independent. Let $u_{2}\left(
k\right) =\tilde{a}\tilde{u}_{2}\left( k\right) ,u_{3}\left( k\right) =%
\tilde{b}\tilde{u}_{2}\left( k\right) $, where $\tilde{u}_{2}\left( k\right)
\in
\mathbb{R}
$. The system (1) is rewritten as%
\begin{equation*}
x\left( k+1\right) =Ax\left( k\right) +u_{1}\left( k\right) B_{1}x\left(
k\right) +\tilde{u}_{2}\left( k\right) \left( \tilde{a}B_{2}+\tilde{b}%
B_{3}\right) x\left( k\right)
\end{equation*}%
which, by the study for the case of $m=2$, is controllable.

Finally, for the necessity, if $A,B_{1},\ldots ,B_{m}$ have a real
eigenvector in common, named $\xi $, then it can be seen that the linear
subspace span$\left\{ \xi \right\} $ is invariant for the system (1). That
is, any state initiated from span$\left\{ \xi \right\} $\ will not leave it,
which makes the system uncontrollable. \ $\square $

$\left. {}\right. $

\textbf{Remark 1.}\textit{\ }Note that Theorem 1 does not require that $%
B_{1},\ldots ,B_{m}$ have rank one or that only one of $u_{1}(k),\ldots
,u_{m}\left( k\right) $ is applied at any one instant; Theorems 6,7 in [14]
and Theorem 1 in [15] are special cases of Theorem 1 when $A$ is a scalar
matrix. The existing results on controllability of system (1) in dimension
two are thus covered.

$\left. {}\right. $

\textbf{Remark 2.}\textit{\ }Compared with the techniques used in [14,15]
which are based on the controllability properties of the single-input
system, the algebraic method applied here makes the controllability proof
simple and straightforward. Furthermore, by using the algebraic method, the
required control inputs which achieve the state transition can be easily
obtained (as shown in Example 1 in Section IV). In particular, the
techniques in [14,15] are not suitable for proving controllability of system
(1), since controllability of the single-input discrete-time bilinear system%
\begin{equation*}
x\left( k+1\right) =Ax\left( k\right) +u\left( k\right) Bx\left( k\right)
\end{equation*}%
is still unknown for general cases even if the system dimension is limited
to two, where $x(k)\in
\mathbb{R}
^{n}$, $A,B\in
\mathbb{R}
^{n\times n}$, and $u(k)\in
\mathbb{R}
$. For the above single-input system without drift ($A=0$), it is
uncontrollable if the system dimension is greater than one.

$\left. {}\right. $

\textbf{Remark 3.}\textit{\ }To verify whether $A,B_{1},\ldots ,B_{m}$ have
a real eigenvector in common is not difficult. We can first consider $A$. If
it does not have a real eigenvector then $A,B_{1},\ldots ,B_{m}$ do not have
a real eigenvector in common. Otherwise, get the real eigenvectors of $A$\
(there exist at most two) and check whether any of them is an eigenvector of
$B_{1},\ldots ,B_{m}$.

$\left. {}\right. $

An example will be provided in Section IV.\textit{\ }We next study
controllability of two-dimensional system (2) and give a necessary and
sufficient condition by using the following lemma.

$\left. {}\right. $

\textbf{Lemma 6.} If $B_{1},B_{2}\in
\mathbb{R}
^{2\times 2}$ are linearly independent and do not have a real eigenvector in
common, then for any nonzero $\xi $ in%
\begin{equation}
\text{span}\left\{ \alpha \right\} \cup \text{span}\left\{ \beta \right\}
\end{equation}%
where $\alpha ,\beta \in
\mathbb{R}
_{\ast }^{2}$ are linearly independent, there exist $a,b\in
\mathbb{R}
$ such that $\left( aB_{1}+bB_{2}\right) \xi $ does not belong to (10) if
and only if $B_{1},B_{2}$ cannot be simultaneously transformed into%
\begin{equation}
PB_{1}P^{-1}=\left[
\begin{array}{cc}
0 & B_{12}^{1} \\
B_{21}^{1} & 0%
\end{array}%
\right] ,\text{ }PB_{2}P^{-1}=\left[
\begin{array}{cc}
0 & B_{12}^{2} \\
B_{21}^{2} & 0%
\end{array}%
\right]
\end{equation}%
where $P\in
\mathbb{R}
^{2\times 2}$ is nonsingular.

$\left. {}\right. $

\textbf{Proof.} The sufficiency proof is similar to that of Lemma 3 and is
thus omitted here.

For the necessity, if $B_{1},B_{2}$ can be simultaneously transformed into
the forms as given in (11), then one can easily verify that the set%
\begin{equation}
\text{span}\left\{ P^{-1}\left[
\begin{array}{cc}
1 & 0%
\end{array}%
\right] ^{T}\right\} \cup \text{span}\left\{ P^{-1}\left[
\begin{array}{cc}
0 & 1%
\end{array}%
\right] ^{T}\right\}
\end{equation}%
is invariant for $\left( aB_{1}+bB_{2}\right) $. \ $\square $

$\left. {}\right. $

\textbf{Theorem 2.} Consider system (2) with $n=2$. If $m=2$, the system is
controllable if and only if $B_{1},B_{2}$ do not have a real eigenvector in
common and cannot be simultaneously transformed into the forms as given in
(11); if $m=3,4$, the system is controllable if and only if $B_{1},\ldots
,B_{m}$ do not have a real eigenvector in common.

$\left. {}\right. $

\textbf{Proof.} Since $n=2$ and $B_{1},\ldots ,B_{m}$ are linearly
independent, we have $m=2,3,4$. Clearly, if $B_{1},\ldots ,B_{m}$ have a
real eigenvector in common, then system (2) is uncontrollable. In the
following, we consider that $B_{1},\ldots ,B_{m}$ do not have a real
eigenvector in common.

For the case of $m=2$, if $B_{1},B_{2}$ can be simultaneously transformed
into the forms as given in (11), then one can verify that the set as given
in (12) is invariant for the system. That is, any state in (12) cannot be
transferred out of (12). Otherwise, for any initial and terminal states $\xi
,\eta $ in $%
\mathbb{R}
_{\ast }^{2}$, either we have%
\begin{equation*}
\left\vert
\begin{array}{cc}
B_{1}\xi  & B_{2}\xi
\end{array}%
\right\vert \neq 0
\end{equation*}%
or we can transfer $\xi $ to a state $\zeta $ which satisfies%
\begin{equation*}
\left\vert
\begin{array}{cc}
B_{1}\zeta  & B_{2}\zeta
\end{array}%
\right\vert \neq 0
\end{equation*}%
by Lemmas 1 and 6. The controllability can be proved as shown in the proof
of Theorem 1.

For the case of $m=3,4$, let $u_{1}\left( k\right) \equiv 1$ then the system
(2) can be regarded as system (1) with two controls or three controls, which
is controllable according to Theorem 1. \ $\square $

$\left. {}\right. $

\section{Near-controllability}

$\left. {}\right. $

If a system is uncontrollable, it is of interest to study the controllable
regions. Near-controllability is established to describe those systems that
are uncontrollable according to the general controllability definition but
have a very large controllable region. This property was first defined and
was demonstrated on two classes of discrete-time bilinear systems [16,17],
and it was then generalized to continuous-time bilinear systems and to
continuous-time and discrete-time nonlinear systems that are not necessarily
bilinear in [18]. Recently, the near-controllability problems were raised in
[19] for discrete-time upper-triangular bilinear systems which are more
general than those considered in [16,17], and necessary conditions and
sufficient conditions for near-controllability were derived. However, the
results in [16,17,19] are for single-input systems only and the study on the
topic of near-controllability is just at the beginning.

$\left. {}\right. $

\textbf{Definition 2 ([16-19]).} A system $\dot{x}\left( t\right) =f\left(
x\left( t\right) ,u\left( t\right) \right) $\ ($x\left( k+1\right) =f\left(
x\left( k\right) ,u\left( k\right) \right) $) is said to be nearly
controllable if, for any $\xi \in
\mathbb{R}
^{n}\left\backslash \mathcal{E}\right. $\ and any $\eta \in
\mathbb{R}
^{n}\left\backslash \mathcal{F}\right. $, there exist piecewise continuous
control $u\left( t\right) $\ and $T>0$\ (a finite control sequence $u\left(
k\right) $, $k=0,1,\ldots ,l-1$, where $l$\ is a positive integer) such that
$\xi $\ can be transferred to $\eta $\ at some $t\in \left( 0,T\right) $\ ($%
k=l$), where $\mathcal{E}$\ and $\mathcal{F}$\ are two sets of zero Lebesgue
measure in $%
\mathbb{R}
^{n}$.

$\left. {}\right. $

It can be seen that if we let $\mathcal{E},\mathcal{F}=\varnothing $, then
the near-controllability definition reduces to the general controllability
definition. Indeed, near-controllability includes the notion of
controllability and can better characterize the properties of control
systems. If we only use \textquotedblleft uncontrollable\textquotedblright\
to describe a system which is not controllable according to the general
controllability definition, then we may miss some valuable properties of it.
In this section, we study the uncontrollable systems (1) and (2) and derive
near-controllability of them.

$\left. {}\right. $

For system (1) in dimension two, from Theorem 1 we know that the system is
uncontrollable if and only if $A,B_{1},\ldots ,B_{m}$ have a real
eigenvector in common. Consider that $A,B_{1},\ldots ,B_{m}$ have a real
eigenvector in common. From the matrix theory we can find a nonsingular
matrix such that all of $A,B_{1},\ldots ,B_{m}$ are transformed into the
following forms%
\begin{equation*}
\left[
\begin{array}{cc}
A_{11} & A_{12} \\
0 & A_{22}%
\end{array}%
\right] ,\text{ }\left[
\begin{array}{cc}
B_{11}^{1} & B_{12}^{1} \\
0 & B_{22}^{1}%
\end{array}%
\right] ,\ldots ,\text{ }\left[
\begin{array}{cc}
B_{11}^{m} & B_{12}^{m} \\
0 & B_{22}^{m}%
\end{array}%
\right] ,
\end{equation*}%
respectively, where $\left[
\begin{array}{cc}
1 & 0%
\end{array}%
\right] ^{T}$ is the common eigenvector and span$\left\{ \left[
\begin{array}{cc}
1 & 0%
\end{array}%
\right] ^{T}\right\} $ is an invariant space for the system (1). Note that $%
A,B_{1},\ldots ,B_{m}$ are linearly independent and $m=2$ or $3$. We have $%
m=2$ and can obtain the following conclusion.

$\left. {}\right. $

\textbf{Theorem 3.} Consider system (1) with $n=2$ and $A,B_{1},\ldots
,B_{m} $ having a real eigenvector in common. Then $m=2$ and there exists a
nonsingular $P\in
\mathbb{R}
^{2\times 2}$ such that the system is transformed into%
\begin{equation}
x\left( k+1\right) =\left( \left[
\begin{array}{cc}
A_{11} & A_{12} \\
0 & A_{22}%
\end{array}%
\right] +u_{1}\left( k\right) \left[
\begin{array}{cc}
B_{11}^{1} & B_{12}^{1} \\
0 & B_{22}^{1}%
\end{array}%
\right] +u_{2}\left( k\right) \left[
\begin{array}{cc}
B_{11}^{2} & B_{12}^{2} \\
0 & B_{22}^{2}%
\end{array}%
\right] \right) x\left( k\right)
\end{equation}%
which is nearly controllable if and only if one of $B_{22}^{1},B_{22}^{2}$
does not vanish.

$\left. {}\right. $

\textbf{Proof.} If $B_{22}^{1}=0,B_{22}^{2}=0$, then%
\begin{equation*}
x_{2}\left( k+1\right) =A_{22}x_{2}\left( k\right)
\end{equation*}%
and system (13) cannot be nearly controllable since both $u_{1}\left(
k\right) ,u_{2}\left( k\right) $ lose the ability on controlling $%
x_{2}\left( k\right) $ such that the system does not have a two-dimensional
controllable region.

If one of $B_{22}^{1},B_{22}^{2}$ does not vanish, assume that it is $%
B_{22}^{1}$ without loss of generality. We can further assume $B_{22}^{2}=0$
since we can let $u_{1}\left( k\right) =\tilde{u}_{1}\left( k\right) -\frac{%
B_{22}^{2}}{B_{22}^{1}}u_{2}\left( k\right) $ where $\tilde{u}_{1}\left(
k\right) \in
\mathbb{R}
$. Then one can verify that for any $\xi $ in $%
\mathbb{R}
^{2}\left\backslash \mathcal{E}\right. $ where%
\begin{equation}
\mathcal{E=}\text{span}\left\{ \left[
\begin{array}{cc}
1 & 0%
\end{array}%
\right] ^{T}\right\} \cup \text{span}\left\{ \left[
\begin{array}{cc}
B_{12}^{2} & -B_{11}^{2}%
\end{array}%
\right] ^{T}\right\}
\end{equation}%
we have%
\begin{equation*}
\left\vert
\begin{array}{cc}
\left[
\begin{array}{cc}
B_{11}^{1} & B_{12}^{1} \\
0 & B_{22}^{1}%
\end{array}%
\right] \xi & \left[
\begin{array}{cc}
B_{11}^{2} & B_{12}^{2} \\
0 & B_{22}^{2}%
\end{array}%
\right] \xi%
\end{array}%
\right\vert \neq 0
\end{equation*}%
and by%
\begin{equation*}
\left[
\begin{array}{c}
u_{1}\left( 0\right) \\
u_{2}\left( 0\right)%
\end{array}%
\right] =\left[
\begin{array}{cc}
\left[
\begin{array}{cc}
B_{11}^{1} & B_{12}^{1} \\
0 & B_{22}^{1}%
\end{array}%
\right] \xi & \left[
\begin{array}{cc}
B_{11}^{2} & B_{12}^{2} \\
0 & B_{22}^{2}%
\end{array}%
\right] \xi%
\end{array}%
\right] ^{-1}\left( \eta -\left[
\begin{array}{cc}
A_{11} & A_{12} \\
0 & A_{22}%
\end{array}%
\right] \xi \right)
\end{equation*}%
the system can be steered from $\xi $ to $\eta $, where $\eta \in
\mathbb{R}
^{2}$. Since $\mathcal{E}$ is a union of two one-dimensional spaces, it has
Lebesgue measure zero in $%
\mathbb{R}
^{2}$. The system is nearly controllable with $\mathcal{E}$ given in (14)
and $\mathcal{F}=\varnothing $. \ $\square $

$\left. {}\right. $

\textbf{Remark 4.} If system (1) in dimension two is neither controllable
nor nearly controllable, it can be transformed into the same form as system
(13) with $B_{22}^{1}=0,B_{22}^{2}=0$, of which the one-dimensional region
span$\left\{ \left[
\begin{array}{cc}
1 & 0%
\end{array}%
\right] ^{T}\right\} $\ is the largest controllable region.

$\left. {}\right. $

For uncontrollable system (2) in dimension two, if $m=2$ then from Theorem 2
$B_{1},B_{2}$ can be simultaneously transformed either into%
\begin{equation}
\left[
\begin{array}{cc}
0 & B_{12}^{1} \\
B_{21}^{1} & 0%
\end{array}%
\right] ,\text{ }\left[
\begin{array}{cc}
0 & B_{12}^{2} \\
B_{21}^{2} & 0%
\end{array}%
\right]
\end{equation}%
or into%
\begin{equation*}
\left[
\begin{array}{cc}
B_{11}^{1} & B_{12}^{1} \\
0 & B_{22}^{1}%
\end{array}%
\right] ,\text{ }\left[
\begin{array}{cc}
B_{11}^{2} & B_{12}^{2} \\
0 & B_{22}^{2}%
\end{array}%
\right] ;
\end{equation*}%
if $m=3$ then from Theorem 2 $B_{1},B_{2},B_{3}$ can be simultaneously
transformed into%
\begin{equation*}
\left[
\begin{array}{cc}
B_{11}^{1} & B_{12}^{1} \\
0 & B_{22}^{1}%
\end{array}%
\right] ,\text{ }\left[
\begin{array}{cc}
B_{11}^{2} & B_{12}^{2} \\
0 & B_{22}^{2}%
\end{array}%
\right] ,\text{ }\left[
\begin{array}{cc}
B_{11}^{3} & B_{12}^{3} \\
0 & B_{22}^{3}%
\end{array}%
\right] .
\end{equation*}%
In addition, for uncontrollable system (2) in dimension two we have $m\neq 4$
since $B_{1},\ldots ,B_{m}$ are linearly independent. We have the following
conclusions on near-controllability of the system (2).

$\left. {}\right. $

\textbf{Theorem 4.} Consider system (2) with $n=2$ and $m=2$. If $%
B_{1},B_{2} $ can be simultaneously transformed into the forms as given in
(15) then the system is nearly controllable; if $B_{1},B_{2}$ have a real
eigenvector in common, then there exists a nonsingular $P\in
\mathbb{R}
^{2\times 2}$ such that the system is transformed into%
\begin{equation}
x\left( k+1\right) =\left( u_{1}\left( k\right) \left[
\begin{array}{cc}
B_{11}^{1} & B_{12}^{1} \\
0 & B_{22}^{1}%
\end{array}%
\right] +u_{2}\left( k\right) \left[
\begin{array}{cc}
B_{11}^{2} & B_{12}^{2} \\
0 & B_{22}^{2}%
\end{array}%
\right] \right) x\left( k\right)
\end{equation}%
which is nearly controllable if and only if one of $B_{22}^{1},B_{22}^{2}$
does not vanish.

$\left. {}\right. $

\textbf{Proof.} If $B_{1},B_{2}$ can be simultaneously transformed into the
forms as given in (15), assume that they are of the forms and one can verify
that for any $\xi $ in $%
\mathbb{R}
^{2}\left\backslash \mathcal{E}\right. $ where%
\begin{equation}
\mathcal{E=}\text{span}\left\{ \left[
\begin{array}{cc}
1 & 0%
\end{array}%
\right] ^{T}\right\} \cup \text{span}\left\{ \left[
\begin{array}{cc}
0 & 1%
\end{array}%
\right] ^{T}\right\}
\end{equation}%
we have%
\begin{equation*}
\left\vert
\begin{array}{cc}
\left[
\begin{array}{cc}
0 & B_{12}^{1} \\
B_{21}^{1} & 0%
\end{array}%
\right] \xi & \left[
\begin{array}{cc}
0 & B_{12}^{2} \\
B_{21}^{2} & 0%
\end{array}%
\right] \xi%
\end{array}%
\right\vert \neq 0
\end{equation*}%
since $B_{1},B_{2}$ are linearly independent. Then by%
\begin{equation*}
\left[
\begin{array}{c}
u_{1}\left( 0\right) \\
u_{2}\left( 0\right)%
\end{array}%
\right] =\left[
\begin{array}{cc}
\left[
\begin{array}{cc}
0 & B_{12}^{1} \\
B_{21}^{1} & 0%
\end{array}%
\right] \xi & \left[
\begin{array}{cc}
0 & B_{12}^{2} \\
B_{21}^{2} & 0%
\end{array}%
\right] \xi%
\end{array}%
\right] ^{-1}\eta
\end{equation*}%
the system can be steered from $\xi $ to $\eta $, where $\eta \in
\mathbb{R}
^{2}$. Furthermore, set (17) is a union of two one-dimensional spaces and
hence has Lebesgue measure zero in $%
\mathbb{R}
^{2}$. The system is nearly controllable with $\mathcal{E}$ given in (17)
and $\mathcal{F}=\varnothing $.

To prove near-controllability of system (16), one can follow from the proof
of Theorem 3. \ $\square $

$\left. {}\right. $

\textbf{Remark 5.} A way to check whether $B_{1},B_{2}$ can be
simultaneously transformed into the forms as given in (15) is first to check
tr$B_{1},$tr$B_{2}$. If one of tr$B_{1},$tr$B_{2}$ does not vanish then $%
B_{1},B_{2}$ cannot be simultaneously transformed into the forms as given in
(15). Otherwise, find the nonsingular matrix such that $B_{1}$ is
transformed into the form as given in (15) and check whether the nonsingular
matrix can make $B_{2}$ be transformed into the same form.

$\left. {}\right. $

\textbf{Theorem 5.} Consider system (2) with $n=2$ and $m=3$. If $%
B_{1},B_{2},B_{3}$ have a real eigenvector in common, then there exists a
nonsingular $P\in
\mathbb{R}
^{2\times 2}$ such that the system is transformed into%
\begin{equation*}
x\left( k+1\right) =\left( u_{1}\left( k\right) \left[
\begin{array}{cc}
B_{11}^{1} & B_{12}^{1} \\
0 & B_{22}^{1}%
\end{array}%
\right] +u_{2}\left( k\right) \left[
\begin{array}{cc}
B_{11}^{2} & B_{12}^{2} \\
0 & B_{22}^{2}%
\end{array}%
\right] +u_{3}\left( k\right) \left[
\begin{array}{cc}
B_{11}^{3} & B_{12}^{3} \\
0 & B_{22}^{3}%
\end{array}%
\right] \right) x\left( k\right)
\end{equation*}%
which is nearly controllable if and only if one of $%
B_{22}^{1},B_{22}^{2},B_{22}^{3}$ does not vanish.

$\left. {}\right. $

\textbf{Proof.} The proof of the necessity is similar to that of Theorem 3.
For the sufficiency, assume that $B_{22}^{2}\neq 0$ without loss of
generality. Letting $u_{1}\left( k\right) =1$ and applying Theorem 3 we can
complete the proof. \ $\square $

$\left. {}\right. $

\textbf{Remark 6.} As discussed in Remark 4, if system (2) in dimension two
is neither controllable nor nearly controllable, then it has at most a
one-dimensional controllable region.

$\left. {}\right. $

\section{Examples}

$\left. {}\right. $

In this section, we give three examples to demonstrate the obtained results
on controllability and near-controllability of systems (1) and (2).

$\left. {}\right. $

\textbf{Example 1.} Consider the system%
\begin{eqnarray}
x\left( k+1\right) &=&\left( A+u_{1}\left( k\right) B_{1}+u_{2}\left(
k\right) B_{2}\right) x\left( k\right)  \notag \\
&=&\left( \left[
\begin{array}{cc}
0 & -1 \\
1 & 0%
\end{array}%
\right] +u_{1}\left( k\right) \left[
\begin{array}{cc}
1 & -1 \\
0 & 2%
\end{array}%
\right] +u_{2}\left( k\right) \left[
\begin{array}{cc}
0 & 0 \\
1 & 0%
\end{array}%
\right] \right) x\left( k\right)
\end{eqnarray}%
with initial state\ $\xi =\left[
\begin{array}{cc}
1 & 1%
\end{array}%
\right] ^{T}$ and terminal state $\eta =\left[
\begin{array}{cc}
-11 & -7%
\end{array}%
\right] ^{T}$, where $x(k)\in
\mathbb{R}
^{2}$ and $u_{1}(k),u_{2}\left( k\right) \in
\mathbb{R}
$.

From (18) $A,B_{1},B_{2}$ are linearly independent and $A$ has none real
eigenvalue which implies $A,B_{1},B_{2}$ do not have a real eigenvector in
common. Therefore, by Theorem 1 system (18) is controllable.

We next find the control inputs to achieve the state transition. Note that%
\begin{equation*}
\left\vert
\begin{array}{cc}
B_{1}\xi & B_{2}\xi%
\end{array}%
\right\vert =0.
\end{equation*}%
Let $u_{1}\left( 0\right) =0,u_{2}\left( 0\right) =0$. We have%
\begin{equation}
x\left( 1\right) =Ax\left( 0\right) =A\xi =\left[
\begin{array}{cc}
-1 & 1%
\end{array}%
\right] ^{T}\triangleq \bar{\xi}  \notag
\end{equation}%
and%
\begin{equation*}
\left\vert
\begin{array}{cc}
B_{1}\bar{\xi} & B_{2}\bar{\xi}%
\end{array}%
\right\vert \neq 0.
\end{equation*}%
From the proof of Theorem 1 by%
\begin{equation*}
\left[
\begin{array}{c}
u_{1}\left( 1\right) \\
u_{2}\left( 1\right)%
\end{array}%
\right] =\left\vert
\begin{array}{cc}
B_{1}\bar{\xi} & B_{2}\bar{\xi}%
\end{array}%
\right\vert ^{-1}\left( \eta -A\bar{\xi}\right) =\left[
\begin{array}{c}
5 \\
16%
\end{array}%
\right]
\end{equation*}%
together with $u_{1}\left( 0\right) ,u_{2}\left( 0\right) $ the system is
steered from $\xi $ to $\eta $.

$\left. {}\right. $

\textbf{Example 2.} Consider the system%
\begin{eqnarray}
x\left( k+1\right) &=&\left( A+u_{1}\left( k\right) B_{1}+u_{2}\left(
k\right) B_{2}\right) x\left( k\right)  \notag \\
&=&\left( \left[
\begin{array}{cc}
5 & 3 \\
-4 & -2%
\end{array}%
\right] +u_{1}\left( k\right) \left[
\begin{array}{cc}
0 & -1 \\
2 & 3%
\end{array}%
\right] +u_{2}\left( k\right) \left[
\begin{array}{cc}
7 & 1 \\
-1 & 5%
\end{array}%
\right] \right) x\left( k\right)
\end{eqnarray}%
where $x(k)\in
\mathbb{R}
^{2}$ and $u_{1}(k),u_{2}\left( k\right) \in
\mathbb{R}
$.

From Remark 2 $A$ has two real eigenvectors $\left[
\begin{array}{cc}
1 & -1%
\end{array}%
\right] ^{T},\left[
\begin{array}{cc}
3 & -4%
\end{array}%
\right] ^{T}$ of which the former is also an eigenvector of$\ B_{1},B_{2}$.
In particular, one can verify that span$\left\{ \left[
\begin{array}{cc}
1 & -1%
\end{array}%
\right] ^{T}\right\} $ is invariant for system (19). By $x\left( k\right)
=P^{-1}\tilde{x}\left( k\right) $ where%
\begin{equation*}
P^{-1}=\left[
\begin{array}{cc}
1 & 0 \\
-1 & 1%
\end{array}%
\right]
\end{equation*}%
we have%
\begin{equation}
\tilde{x}\left( k+1\right) =\left( \left[
\begin{array}{cc}
2 & 3 \\
0 & 1%
\end{array}%
\right] +u_{1}\left( k\right) \left[
\begin{array}{cc}
1 & -1 \\
0 & 2%
\end{array}%
\right] +u_{2}\left( k\right) \left[
\begin{array}{cc}
6 & 1 \\
0 & 6%
\end{array}%
\right] \right) \tilde{x}\left( k\right) .
\end{equation}%
According to Theorem 3, system (20) is nearly controllable with%
\begin{equation*}
\mathcal{E=}\text{span}\left\{ \left[
\begin{array}{cc}
1 & 0%
\end{array}%
\right] ^{T}\right\} \cup \text{span}\left\{ \left[
\begin{array}{cc}
1 & -6%
\end{array}%
\right] ^{T}\right\} ,\text{ }\mathcal{F}=\varnothing .
\end{equation*}%
Thus, system (19) is nearly controllable with%
\begin{eqnarray*}
\mathcal{E} &\mathcal{=}&\text{span}\left\{ P^{-1}\left[
\begin{array}{cc}
1 & 0%
\end{array}%
\right] ^{T}\right\} \cup \text{span}\left\{ P^{-1}\left[
\begin{array}{cc}
1 & -6%
\end{array}%
\right] ^{T}\right\} =\text{span}\left\{ \left[
\begin{array}{cc}
1 & -1%
\end{array}%
\right] ^{T}\right\} \cup \text{span}\left\{ \left[
\begin{array}{cc}
1 & -7%
\end{array}%
\right] ^{T}\right\} , \\
\mathcal{F} &=&\varnothing .
\end{eqnarray*}

$\left. {}\right. $

\textbf{Example 3.} Consider the system%
\begin{eqnarray}
x\left( k+1\right) &=&\left( u_{1}\left( k\right) B_{1}+u_{2}\left( k\right)
B_{2}\right) x\left( k\right)  \notag \\
&=&\left( u_{1}\left( k\right) \left[
\begin{array}{cc}
-1 & 0 \\
3 & 1%
\end{array}%
\right] +u_{2}\left( k\right) \left[
\begin{array}{cc}
4 & 3 \\
-6 & -4%
\end{array}%
\right] \right) x\left( k\right)
\end{eqnarray}%
where $x(k)\in
\mathbb{R}
^{2}$ and $u_{1}(k),u_{2}\left( k\right) \in
\mathbb{R}
$.

From Remark 3 we note that tr$B_{1}=0,$ tr$B_{2}=0$. By $\tilde{x}\left(
k\right) =Px\left( k\right) $ where%
\begin{equation*}
P=\left[
\begin{array}{cc}
2 & 1 \\
1 & 1%
\end{array}%
\right]
\end{equation*}%
we have%
\begin{equation}
\tilde{x}\left( k+1\right) =\left( u_{1}\left( k\right) \left[
\begin{array}{cc}
0 & 1 \\
1 & 0%
\end{array}%
\right] +u_{2}\left( k\right) \left[
\begin{array}{cc}
0 & 2 \\
-1 & 0%
\end{array}%
\right] \right) \tilde{x}\left( k\right) .
\end{equation}%
According to Theorem 4, system (22) is nearly controllable with%
\begin{equation*}
\mathcal{E=}\text{span}\left\{ \left[
\begin{array}{cc}
1 & 0%
\end{array}%
\right] ^{T}\right\} \cup \text{span}\left\{ \left[
\begin{array}{cc}
0 & 1%
\end{array}%
\right] ^{T}\right\} ,\text{ }\mathcal{F}=\varnothing .
\end{equation*}%
Thus, system (21) is nearly controllable with%
\begin{eqnarray*}
\mathcal{E} &\mathcal{=}&\text{span}\left\{ P^{-1}\left[
\begin{array}{cc}
1 & 0%
\end{array}%
\right] ^{T}\right\} \cup \text{span}\left\{ P^{-1}\left[
\begin{array}{cc}
0 & 1%
\end{array}%
\right] ^{T}\right\} =\text{span}\left\{ \left[
\begin{array}{cc}
1 & -1%
\end{array}%
\right] ^{T}\right\} \cup \text{span}\left\{ \left[
\begin{array}{cc}
-1 & 2%
\end{array}%
\right] ^{T}\right\} , \\
\mathcal{F} &=&\varnothing .
\end{eqnarray*}

$\left. {}\right. $

\section{Conclusions}

$\left. {}\right. $

Although the controllability problem of two-dimensional single-input
discrete-time bilinear systems remains unsolved, in this paper, the
controllability problems of two-dimensional multi-input discrete-time
bilinear systems with and without drift are completely solved. Necessary and
sufficient conditions for controllability are obtained by using an algebraic
method. For the uncontrollable systems, near-controllability is studied and
necessary and sufficient conditions for the systems to be nearly
controllable are also presented. Finally, examples are provided to
demonstrate the conceptions and results of the paper. Future work should
consider the controllability and near-controllability problems of
single-input as well as multi-input discrete-time bilinear systems in high
dimensions.

$\left. {}\right. $

\section{Appendix}

$\left. {}\right. $

\textbf{Proof of Lemma 3.} Note that set (6) only contains the vectors that
are linearly dependent on $\alpha $ or on $\beta $. Without loss of
generality, we can let $\xi =\alpha $ or $\xi =\beta $. If for any $a,b\in
\mathbb{R}
$ $\left( A+aB_{1}+bB_{2}\right) \xi $ is in (6), then for a given pair of $%
a,b$ we have the following four cases:%
\begin{eqnarray*}
\left( A+aB_{1}+bB_{2}\right) \left[
\begin{array}{cc}
\alpha & \beta%
\end{array}%
\right] &=&\left[
\begin{array}{cc}
\alpha & \beta%
\end{array}%
\right] \left[
\begin{array}{cc}
p & 0 \\
0 & q%
\end{array}%
\right] ; \\
\left( A+aB_{1}+bB_{2}\right) \left[
\begin{array}{cc}
\alpha & \beta%
\end{array}%
\right] &=&\left[
\begin{array}{cc}
\alpha & \beta%
\end{array}%
\right] \left[
\begin{array}{cc}
p & q \\
0 & 0%
\end{array}%
\right] ; \\
\left( A+aB_{1}+bB_{2}\right) \left[
\begin{array}{cc}
\alpha & \beta%
\end{array}%
\right] &=&\left[
\begin{array}{cc}
\alpha & \beta%
\end{array}%
\right] \left[
\begin{array}{cc}
0 & q \\
p & 0%
\end{array}%
\right] ; \\
\left( A+aB_{1}+bB_{2}\right) \left[
\begin{array}{cc}
\alpha & \beta%
\end{array}%
\right] &=&\left[
\begin{array}{cc}
\alpha & \beta%
\end{array}%
\right] \left[
\begin{array}{cc}
0 & 0 \\
p & q%
\end{array}%
\right]
\end{eqnarray*}%
where $p,q$ are some real numbers. Since $\left[
\begin{array}{cc}
\alpha & \beta%
\end{array}%
\right] $ is nonsingular and a nonsingular transformation does not affect
the proof\footnote{%
For instance, for the first case we have
\par
\begin{align*}
\left( A+aB_{1}+bB_{2}\right) \left[
\begin{array}{cc}
\alpha & \beta%
\end{array}%
\right] & =\left[
\begin{array}{cc}
\alpha & \beta%
\end{array}%
\right] \left[
\begin{array}{cc}
p & 0 \\
0 & q%
\end{array}%
\right] \\
& \Leftrightarrow \left[
\begin{array}{cc}
\alpha & \beta%
\end{array}%
\right] ^{-1}\left( A+aB_{1}+bB_{2}\right) \left[
\begin{array}{cc}
\alpha & \beta%
\end{array}%
\right] =\left[
\begin{array}{cc}
p & 0 \\
0 & q%
\end{array}%
\right] \\
& \Leftrightarrow \left( \left[
\begin{array}{cc}
\alpha & \beta%
\end{array}%
\right] ^{-1}A\left[
\begin{array}{cc}
\alpha & \beta%
\end{array}%
\right] +a\left[
\begin{array}{cc}
\alpha & \beta%
\end{array}%
\right] ^{-1}B_{1}\left[
\begin{array}{cc}
\alpha & \beta%
\end{array}%
\right] +b\left[
\begin{array}{cc}
\alpha & \beta%
\end{array}%
\right] ^{-1}B_{2}\left[
\begin{array}{cc}
\alpha & \beta%
\end{array}%
\right] \right) =\left[
\begin{array}{cc}
p & 0 \\
0 & q%
\end{array}%
\right] .
\end{align*}%
\par
Then, the new \textquotedblleft $\left[
\begin{array}{cc}
\alpha & \beta%
\end{array}%
\right] $\textquotedblright\ for Case 1 after transformation is $I$.
\par
{}
\par
{}}, we may let%
\begin{equation*}
\left[
\begin{array}{cc}
\alpha & \beta%
\end{array}%
\right] =\left[
\begin{array}{cc}
1 & 0 \\
0 & 1%
\end{array}%
\right] =I
\end{equation*}%
without loss of generality. As a result, we study the following cases%
\begin{align*}
1.\text{ }A\alpha & \in \text{span}\left\{ \alpha \right\} ,\text{ }A\beta
\in \text{span}\left\{ \beta \right\} ; \\
2.\text{ }A\alpha ,A\beta & \in \text{span}\left\{ \alpha \right\} ; \\
3.\text{ }A\alpha & \in \text{span}\left\{ \beta \right\} ,\text{ }A\beta
\in \text{span}\left\{ \alpha \right\} ; \\
4.\text{ }A\alpha ,A\beta & \in \text{span}\left\{ \beta \right\} .
\end{align*}

For Case 1, let $a\neq 0,b=0$ and $a\neq 0,b\neq 0$ respectively, then it is
easy to see that both $\alpha ,\beta $ are common eigenvectors of $%
A,B_{1},B_{2}$, which leads to a contradiction.

For Case 2, we can write%
\begin{equation*}
A=\left[
\begin{array}{cc}
A_{11} & A_{12} \\
0 & 0%
\end{array}%
\right] .
\end{equation*}%
If $\left( A+aB_{1}\right) \alpha \in $span$\left\{ \alpha \right\} $ for
some nonzero $a$, we can write%
\begin{equation*}
B_{1}=\left[
\begin{array}{cc}
B_{11}^{1} & B_{12}^{1} \\
0 & B_{22}^{1}%
\end{array}%
\right] .
\end{equation*}%
Write%
\begin{equation}
B_{2}=\left[
\begin{array}{cc}
B_{11}^{2} & B_{12}^{2} \\
B_{21}^{2} & B_{22}^{2}%
\end{array}%
\right] .
\end{equation}%
For nonzero $b$ we have $\left( A+aB_{1}+bB_{2}\right) \alpha \notin $span$%
\left\{ \alpha \right\} $ (otherwise it can be deduced that $B_{21}^{2}=0$
and $\alpha $ is the common eigenvector of $A,B_{1},B_{2}$) and hence $%
\left( A+aB_{1}+bB_{2}\right) \alpha \in $span$\left\{ \beta \right\} $.
Letting $a\neq 0$ while $b$ vary and $b\neq 0$ while $a$ vary, respectively,
we can deduce $A_{11}=0,B_{11}^{1}=0,B_{11}^{2}=0$ (thus, $A_{12}\neq 0$
otherwise $A=0$ which contradicts that $A,B_{1},B_{2}$ are linearly
independent). Then%
\begin{equation*}
\left( A+aB_{1}+bB_{2}\right) \beta =\left[
\begin{array}{c}
A_{12}+aB_{12}^{1}+bB_{12}^{2} \\
aB_{22}^{1}+bB_{22}^{2}%
\end{array}%
\right]
\end{equation*}%
which must belong to span$\left\{ \alpha \right\} $ since $A_{12}\neq 0$ and
hence we have $B_{22}^{1}=0,B_{22}^{2}=0$. This implies that $A,B_{1}$ are
linearly dependent which contradicts that $A,B_{1},B_{2}$ are linearly
independent.

Therefore, $\left( A+aB_{1}\right) \alpha \in $span$\left\{ \beta \right\} $
for any nonzero $a$ such that $B_{1}$ can only be written as%
\begin{equation*}
B_{1}=\left[
\begin{array}{cc}
0 & B_{12}^{1} \\
B_{21}^{1} & B_{22}^{1}%
\end{array}%
\right]
\end{equation*}%
and $A_{11}=0$ (thus, $A_{12}\neq 0$). Write $B_{2}$ as given in (7). Then%
\begin{equation*}
\left( A+aB_{1}+bB_{2}\right) \alpha =\left[
\begin{array}{c}
bB_{11}^{2} \\
aB_{21}^{1}+bB_{21}^{2}%
\end{array}%
\right]
\end{equation*}%
which belongs either to span$\left\{ \alpha \right\} $ or to span$\left\{
\beta \right\} $ for any $a,b$. This implies that either $%
B_{21}^{1}=0,B_{21}^{2}=0$ or $B_{11}^{2}=0$. For the former case we have
that $\alpha $ is the common eigenvector of $A,B_{1},B_{2}$; for the latter
case, since $A_{12}\neq 0$, we can only have $\left( A+aB_{1}+bB_{2}\right)
\beta \in $span$\left\{ \alpha \right\} $ for any $a,b$ which implies that $%
B_{22}^{1}=0,B_{22}^{2}=0$ and then%
\begin{equation*}
A=\left[
\begin{array}{cc}
0 & A_{12} \\
0 & 0%
\end{array}%
\right] ,\text{ }B_{1}=\left[
\begin{array}{cc}
0 & B_{12}^{1} \\
B_{21}^{1} & 0%
\end{array}%
\right] ,\text{ }B_{2}=\left[
\begin{array}{cc}
0 & B_{12}^{2} \\
B_{21}^{2} & 0%
\end{array}%
\right]
\end{equation*}%
must be linearly independent. Therefore, for Case 2 the contradiction is
made.

For Case 3, we can write%
\begin{equation*}
A=\left[
\begin{array}{cc}
0 & A_{12} \\
A_{21} & 0%
\end{array}%
\right] .
\end{equation*}%
If $\left( A+aB_{1}\right) \alpha \in $span$\left\{ \alpha \right\} $ for
some $a$, we have $A_{21}=0$ (then $A_{12}\neq 0$) and can write%
\begin{equation*}
B_{1}=\left[
\begin{array}{cc}
B_{11}^{1} & B_{12}^{1} \\
0 & B_{22}^{1}%
\end{array}%
\right] .
\end{equation*}%
Write $B_{2}$ as given in (7). For nonzero $b$ we have $\left(
A+aB_{1}+bB_{2}\right) \alpha \notin $span$\left\{ \alpha \right\} $
(otherwise $\alpha $ is the common eigenvector of $A,B_{1},B_{2}$) and hence
$\left( A+aB_{1}+bB_{2}\right) \alpha \in $span$\left\{ \beta \right\} $.
Letting $a\neq 0$ while $b$ vary and $b\neq 0$ while $a$ vary, respectively,
we can deduce $B_{11}^{1}=0,B_{11}^{2}=0$. Then%
\begin{equation*}
\left( A+aB_{1}+bB_{2}\right) \beta =\left[
\begin{array}{c}
A_{12}+aB_{12}^{1}+bB_{12}^{2} \\
aB_{22}^{1}+bB_{22}^{2}%
\end{array}%
\right]
\end{equation*}%
which must belong to span$\left\{ \alpha \right\} $ since $A_{12}\neq 0$ and
hence we have $B_{22}^{1}=0,B_{22}^{2}=0$. This implies that $A,B_{1}$ are
linearly dependent which contradicts that $A,B_{1},B_{2}$ are linearly
independent.

Therefore, $\left( A+aB_{1}\right) \alpha \in $span$\left\{ \beta \right\} $
for any nonzero $a$ such that $B_{1}$ can only be written as%
\begin{equation*}
B_{1}=\left[
\begin{array}{cc}
0 & B_{12}^{1} \\
B_{21}^{1} & B_{22}^{1}%
\end{array}%
\right] .
\end{equation*}%
Still write $B_{2}$ as given in (23). Then%
\begin{equation*}
\left( A+aB_{1}+bB_{2}\right) \alpha =\left[
\begin{array}{c}
bB_{11}^{2} \\
A_{21}+aB_{21}^{1}+bB_{21}^{2}%
\end{array}%
\right]
\end{equation*}%
which belongs either to span$\left\{ \alpha \right\} $ or to span$\left\{
\beta \right\} $ for any $a,b$. This implies that either $%
A_{21}=0,B_{21}^{1}=0,B_{21}^{2}=0$ or $B_{11}^{2}=0$. For the former case
we have that $\alpha $ is the common eigenvector of $A,B_{1},B_{2}$; for the
latter case we can deduce $A_{21}=0,B_{21}^{1}=0,B_{21}^{2}=0$ or $%
B_{22}^{1}=0,B_{22}^{2}=0$, either of which will lead to a contradiction.

Finally, for Case 4, it is a symmetry to Case 2. \ $\square $

$\left. {}\right. $

\end{document}